\documentclass[english,aps,prstper,reprint,showpacs,titlepage, longbibliography]{revtex4-2}   
\usepackage[utf8]{inputenc}
\usepackage[T1]{fontenc}	
\usepackage{geometry}		
\usepackage{graphicx}
\usepackage[above,below]{placeins}	
\usepackage{times}

\usepackage{hyperref}  
\hypersetup{colorlinks=true,urlcolor=blue,citecolor=blue,linkcolor=blue}   
\usepackage{enumitem}          
 \setlist{nosep}                 

\usepackage{tikz}
\usepackage{graphicx}
\usepackage{xcolor, soul}
\sethlcolor{black}
\usepackage{array}
\newcolumntype{C}[1]{>{\centering\arraybackslash}m{#1}}

\begin{document}

\begin{titlepage}

\title{Industry Perspectives on Projected Quantum Workforce Needs}
\author{Shams El-Adawy\textsuperscript{1,2}}

\author{A.R. Pi\~na\textsuperscript{3}}

\author{Benjamin M. Zwickl\textsuperscript{3}}

\author{H. J. Lewandowski\textsuperscript{1,2}}

\affiliation{\textsuperscript{1}JILA, National Institute of Standards and Technology and the University of Colorado, Boulder, Colorado 80309, USA}
\affiliation{\textsuperscript{2}Department of Physics, University of Colorado, Boulder, Colorado 80309, USA }
\affiliation{\textsuperscript{3}School of Physics and Astronomy, Rochester Institute of Technology, Rochester, NY 14623, USA}

\begin{abstract}
As more physics educators are developing courses and programs to prepare students for careers in quantum information science, understanding the quantum industry's future workforce needs has become increasingly important. As part of ongoing efforts to understand the knowledge and skills needed for various job roles, we interviewed quantum industry professionals in managerial positions about workforce needs. Through thematic analysis, we identify two broad themes about projected needs. First, managers anticipate a need for a range of educational levels from bachelors to PhDs in physics, engineering, and computer science to fill the needs of roles spanning manufacturing to innovation. Second, managers anticipate an increased need for individuals who can apply quantum information science knowledge across fields. These results provide insights for physics educators about course and program development: continued investment in quantum information science education at all levels is valuable, and greater emphasis should be placed on applications of quantum science.  \\
 \clearpage
\end{abstract}

\maketitle
\end{titlepage}

\section{Introduction}

Quantum Information Science and Engineering (QISE) is an interdisciplinary field that is at the core of advancements in quantum computing, networking and communication, and sensing. As QISE research and development progresses, it holds the potential to transform industries by enabling new quantum technologies, from secure communication to faster computing systems. This promising growth has catalyzed significant investment, with several key pieces of legislation across the globe fueling both QISE research and education initiatives \cite{NQI,riedel2019europe, knight2019uk, NSP, CHIPS}. In the United States, the National Quantum Initiative Act \cite{NQI}, the Quantum Information Science and Technology Workforce Development National Strategic Plan \cite{NSP}, and the CHIPS and Science Act \cite{CHIPS} have prioritized QISE by supporting research in QISE and the development and assessment of QISE education. 

As the need for a well-prepared quantum workforce grows alongside technological advancements, it becomes increasingly important to better understand how to prepare the individuals that will be driving and supporting these innovations. Research to-date highlights the importance of preparing students at multiple educational levels to support a range of roles in the quantum industry \cite{fox2020preparing,hughes2022assessing, kaur2022defining, douglass2023quantum, goorney2025quantum}. In particular, Fox \textit{et al.} \cite{fox2020preparing} point to the need for both advanced quantum expertise at the doctoral level and strong technical, software development, and engineering skills at the undergraduate level, while Hughes \textit{et al.} \cite{hughes2022assessing} reinforce that companies are hiring across all degree levels to support quantum technology development and deployment. 

Additional research has stressed the value of experiential learning in helping students connect academic training with workforce expectations \cite{aiello2021achieving, li2024developing, qedc2025experiential}. For example, Aiello \textit{et al.} explain how internships, capstone projects, and/or laboratory-based courses not only help students get a sense of what a job in QISE looks like, but also showcase practical skills sought out by industry \cite{aiello2021achieving}. Additionally, Li \textit{et al.} demonstrate the importance of experiential learning in supporting the development of quantum literacy \cite{li2024developing}. 

In parallel, a growing body of literature has examined various dimensions of QISE education. Studies have examined the landscape of quantum and QISE courses and programs across institutions \cite{cervantes2021overview,buzzell2025quantum,pina2025landscape}. Other work has focused on creating QISE-specific curricula \cite{devore2016development, economou2020teaching,salehi2021computer} and explored the opportunities and challenges educators face in designing and developing QISE courses and programs \cite{quantumeducators}. Collectively, these studies provide insights on ongoing educational efforts and where further efforts are needed to support sustained course and program growth.  

While this body of work on QISE education and workforce development has advanced our understanding of how to prepare students for existing quantum roles, it remains largely focused on current workforce needs. In other words, these studies offer a valuable, but time-bound snapshot of how academia can meet present-day industry expectations. What remains less well explored are the ways in which industry expectations of educational requirements and job roles might shift in the coming years as the quantum sector evolves. This gap limits the ability of educators to design proactive QISE courses and programs aimed at building a sustainable quantum workforce pipeline. In particular, since it can take a few years to design new academic programs, designing with future needs in mind can be quite beneficial. 

Thus, this paper aims to address this need by exploring broader trends expected to shape the quantum workforce.  We do not provide a detailed analysis of the specific knowledge, skills, and abilities needed for current job roles in the industry, as that is the focus of separate ongoing work, which will allow us to compare our findings with Greinert \textit{et al.}'s similar work in the European context \cite{greinert2024advancing}. Instead, this paper addresses the following research question (RQ): \textbf{How do quantum industry managers anticipate  workforce trends evolving over the next five years in terms of educational requirements and job roles?}

\section{Methodology}

\subsection{Data collection}

Our data consisted of research interviews with quantum industry professionals. Our recruitment process involved multiple strategies. We began by identifying quantum companies affiliated with the Quantum Economic Development Consortium (QED-C)  \cite{qedc_homepage}, a consortium that consolidates and supports various efforts from national stakeholders in the quantum ecosystem. The QED-C provided some contact information for professionals at affiliated quantum companies.  Simultaneously, we leveraged our institutional networks, including alumni in the quantum industry, to identify additional participants. We also used a snowball sampling approach \cite{parker2019snowball} by asking each interviewee to refer colleagues within their organization to participate in our research interviews, which allowed us to have data from a variety of roles within a single company. 

In total, we conducted 34 interviews with employees and managers from December 2024 to April 2025. The manager interviews focused on company’s role within the quantum industry, skills and knowledge of specific positions, and interviewee's outlook on the quantum industry. In contrast, employee interviewees focused on the interviewee's position, the specific tasks they perform, and the associated knowledge and skills needed to do those tasks. To answer the RQ in this paper, we focus on only the managers' interviews because we explicitly asked them about their outlook on the industry, which provides a high-level overview of anticipated needs.

\begin{figure*}
    \centering
    \includegraphics[width=0.85\linewidth]{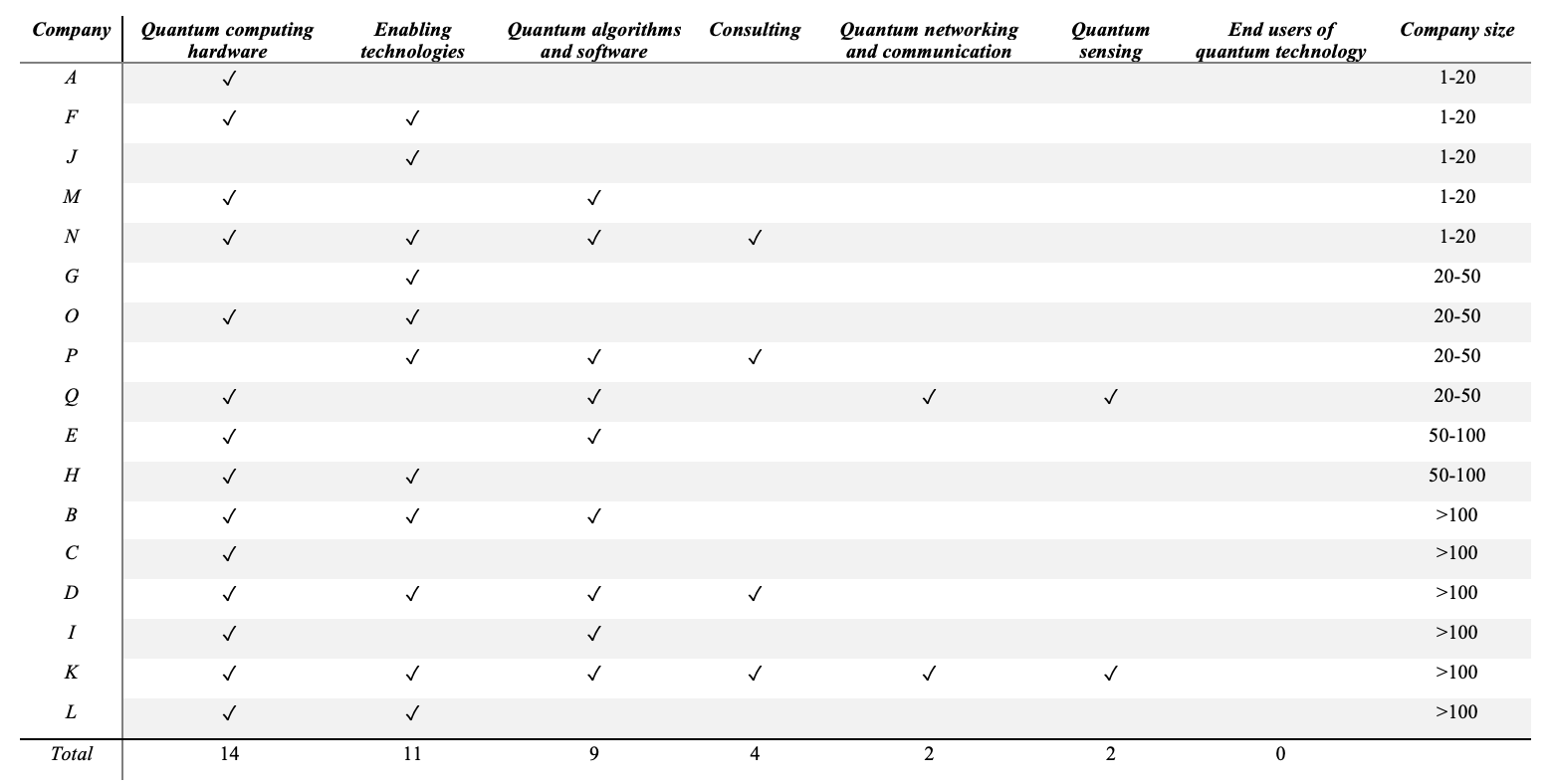}
    \caption{Company categorization of activities and company size based on the 20 quantum industry manager interviews}
    \label{fig:companycategorization}
\end{figure*}

We interviewed 20 professionals in managerial roles at 17 companies. These managers held a range of roles, including CEOs, founders, talent acquisition managers, and senior scientists. The types of activities their companies engaged in are listed in Figure \ref{fig:companycategorization}. We adopt definitions consistent with prior literature \cite{fox2020preparing} to describe the companies, and we let interviewees select the classification that best fit their company. Briefly summarized the categories of companies are:
\begin{itemize}
    \item \textbf{Quantum computing hardware} companies build quantum computers along with necessary software to operate and simulate them.
    \item \textbf{Quantum algorithms and software} companies apply quantum computation to solve real-world problems and develop new algorithms for quantum computers.
    \item \textbf{Enabling technologies} companies are producing customized hardware used in quantum sensors, networking and communication, or computing hardware.
    \item \textbf{Quantum networking and communication} companies focus on creating technologies for quantum-key distribution or hardware for transmitting entangled quantum states.
    \item \textbf{Quantum sensing} companies are developing precise sensors by leveraging controlled quantum states for commercially viable applications.
    \item \textbf{Consulting} companies provide expertise and strategic advice on adopting or integrating quantum technologies and navigating the quantum ecosystem.
\end{itemize}
We also identified company sizes based on interviewees' responses, using the following ranges: 1-20, 20-50, 50-100 or more than 100 employees. 

\subsection{Data analysis}
To answer our RQ, we focused on interviewees' responses to these two interview questions about the quantum industry as a whole:
\begin{itemize}
    \item What types of degrees, fields, certifications and experience are most likely to be needed for developing the quantum industry in the next five years?
    \item What specific jobs are most likely to be needed for developing the quantum industry in the next five years?
\end{itemize}
Thematic analysis \cite{clarke2014thematic} started by reading all responses to both questions to code pieces of the data related to qualifications, experience, and job roles. Initial codes were generated, which included  terms such as “PhD for innovation”, “manufacturing roles”, and “application area”. These codes were then grouped to generate broader themes about future needs of the quantum industry, which resulted in two distinct themes. To ensure consistency, a second researcher independently coded a subset of the data (20 quotes from across the interviews) using the identified themes. There was a 95\% agreement between coders on this subset, which demonstrated that the themes captured the patterns in the data clearly and comprehensively. We report percent agreement instead of Cohen’s kappa due to the small subset of the dataset used for this analysis and the presence of only two major themes. 

There are a few limitations in our data collection that limit the scope of interpretation of our results. First, the sample of managers is not representative of the full range of company types within the quantum industry. For example, as demonstrated in Figure \ref{fig:companycategorization},  end users of quantum technology companies are not part of our sample yet, even though one of the themes in our findings is that positions may filled by many employees working on those application-focused companies. Second, the data analyzed in this paper represents a small subset of a larger dataset currently being collected. As additional interviews are conducted and analyzed, further nuances and perspectives may emerge that are not captured here.

\section{Results}

Two themes emerge from our analysis. Theme 1, consistent with previous literature \cite{fox2020preparing,hughes2022assessing, kaur2022defining}, highlights a continued projected need for a range of higher educational levels in physics, engineering, and computer science to drive manufacturing and innovation as the industry scales.
Theme 2, a newer insight, points to an anticipated need for more individuals who can apply quantum information science knowledge in various sectors.

\subsection{Theme 1: Educational Requirements}
The need for a workforce spanning a range of educational levels to support the scaling of the quantum industry is a theme that was identified in 17 out of 20 of the quantum industry manager interviews. Quantum industry managers emphasized that PhD-level talent will continue to be essential for QISE, particularly for innovation in foundational research. One manager at company I articulates this idea clearly, saying:
\begin{quote}
    \textit{There's scaling up and then scaling out. The scaling-up is the PhDs. We still need a ton of those and we still need to innovate.}
\end{quote}
The continued need for PhD-level employees is echoed across company types and sizes, and underscores how advanced expertise remains a cornerstone of industry growth. 

However, as quantum technologies move beyond research and development toward implementation, managers also anticipate a growing reliance on individuals from across educational levels. Individuals with bachelor’s and master’s degrees are seen as critical to supporting this transition. One manager at company M captures this shift:
\begin{quote}
    \textit{It's primarily going to be driven by, either PhDs or bachelor level students. It's that split dichotomy of you're going to have the folks who need to engineer what needs to be done and then the people who execute on it. In particular, technician or maybe early engineer stage is where you see bachelors going in.}
\end{quote}
This manager outlines the potential emergence of a division of labor with the quantum workforce: PhDs leading design and innovation, and bachelor's professionals supporting execution and maintenance. 

In parallel, managers also highlight the continued relevance of foundational disciplines. Physics, engineering, and computer science are still indispensable to quantum technology development. As one manager at company K explains:
\begin{quote}
   \textit{ If you look at cryogenic test systems, they need mechanical engineers still. They need thermal people. If you look at the quantum measurement side, they still need more physicists. If you look at building the hardware, going back to the test, you still need hardware engineers. I would say, the classics are still necessary. Everything from physics to mechanical to electrical to computer science, it's all needed.}
\end{quote}
This need is particularly emphasized by managers at larger companies, those with more than 100 employees working on quantum, which reinforces the idea that as teams grow,  a broader range of expertise across foundational disciplines becomes increasingly necessary to build quantum systems.

Looking ahead, to support the transition from experimental research to manufacturing and commercial deployment, managers foresee a growing need for technicians to manage the assembly, production, and maintenance of systems. A manager at company G illustrate the hands-on nature of these roles by describing typical  technician tasks in their company to include fiber splicing, cleaning and assembling optical components, and building optical setups. They went on to describe the broader category of roles as:
\begin{quote}
    \textit{As we transition from laboratory proof-of-concepts to actually deployed manufactured units, the fraction of folks that are on the manufacturing side are going to grow.}
\end{quote}
This projected need for more individuals in manufacturing roles, echoed by other managers in quantum hardware and enabling technologies companies, signals the expansion of quantum workforce roles in manufacturing, reinforcing the need for a broad and layered talent pipeline. 

\subsection{Theme 2: Application-Focused Roles}
The second major theme that emerged from these interviews, discussed by 19 out of 20 managers, is  the growing need for professionals who can operate at the intersection of quantum science and applied fields. As the quantum industry matures, there is an anticipated shift from an emphasis on fundamental research toward practical implementation. Thus, there is an expectation that there is a need for professionals who not only understand quantum principles, but also can bridge these concepts into practical applications.

One manager at company D describes the importance of applied quantum work:
\begin{quote}
    \textit{I think as you see this focus shift away from hardware-oriented tasks and more towards application area, you can see the [quantum] industry evolve from applied research to actually applicable to [other] industry.}
\end{quote}
This manager suggests a transition from building the tools to using them effectively. Echoing on this idea, there is an emphasis on multidisciplinary training, which is seen as essential for connecting quantum theory to real world challenges. For example, another manager at the same company D says:
\begin{quote}
    \textit{I would say, if you're studying computer science or you're studying data science, if you're studying specific domain areas that quantum computing shows promise in chemistry, finance, AI, optimization, if you're studying in those areas, if you're thinking about quantum computation and how to actually connect those dots  between those areas, those are all super valuable skills. I think that's, I don't think we have a word to describe all those, the combination of things, but those are the sorts of areas that I think would be valuable for students to focus on.}
\end{quote}
This manager captures the emerging need for cross-domain fluency. 

As companies look beyond technical development towards integration into commercial sectors, there is an anticipated need for professionals not only in technical roles, but also in positions that support the industry’s scaling. As quantum technologies move into the public and commercial domains, the ability to communicate about the impact of these technologies becomes increasingly important. As one manager at company O explains:
\begin{quote}
    \textit{I think moving forward you're going to see a lot of more soft-skill-based degrees coming into quantum, like business bachelors and sales and marketing, and  digital creators. Because we are going from lab to industry, and part of doing so is communicating with a broader audience.}
\end{quote}
Echoed by managers across company types and sizes, this quote signals a potential expansion of the talent pipeline beyond traditional scientists and engineers to include those who can explain and advocate for quantum technologies in the broader market.

As a result, managers anticipate the emergence of integration roles that can bridge technical depth with broad application knowledge.  A manager at company Q summarizes this outlook:
\begin{quote}
    \textit{I would foresee us having some integration role in the future. I don't know if everybody will, but I think that's where we really need folks with that multidisciplinary knowledge of those other subjects in engineering to be able to translate that to quantum programs.}
\end{quote}
This insight reflects the projected need for professionals who are fluent in both quantum and non-quantum domains. 

\section{Discussion}
Our analysis identifies two themes regarding workforce needs in the quantum industry. The two themes are based on industry leaders projecting to move from research and development to commercial deployment in the near future. First, managers anticipate the continued necessity of PhD-level expertise to support innovation, while also projecting a growing need for individuals with bachelor and master’s degree in physics, engineering, and computer science to support manufacturing. Second, there is an expectation that future quantum professionals should be able to apply QISE knowledge across various domains. 

Theme 1 underscores the value of building capacity across the entire educational spectrum. Institutions are already starting to invest in QISE bachelor's and master's level courses and programs \cite{perron2021quantum, economou2022hello, asfaw2022building,blanchette2024design,goorney2024quantum}, and continued investment is essential to develop a layered talent pipeline to support the quantum industry's growth.
Theme 2 points to the need for academic courses and programs to integrate application-oriented QISE content into existing pathways. For example, institutions could consider offering at least one course in quantum applications for all STEM majors. Moreover, upskilling opportunities such as certificates  can help individuals who are in adjacent industries transition into quantum-related roles. 

However, a nuance emerged in our interviews that complicates how we approach educational planning for the QISE workforce.  In our interviews, we ask managers about current roles and anticipated roles. Their responses about current roles (the focus of ongoing separate work) tended to emphasize positions requiring advanced quantum expertise. Yet, when asked to project future workforce needs (the focus of this paper), these same managers describe a broader workforce, one that increasingly includes bachelor-level employees. This difference may reflect hiring patterns that still favor advanced expertise, even as managers maintain an arguably optimistic outlook on the growth and widespread adoption of applied quantum technologies. This difference poses challenges for educational planning, as preparation must span a wide range of roles from advanced quantum expert roles, to multi-disciplinary experts roles, and manufacturing roles. Our ongoing data collection and analysis for this project may provide further insight into how to approach planning for QISE education given how the current and future needs of the industry differ.

Nevertheless, to support the field's development, coordinated efforts among stakeholders, including industry, academia, and government, are essential to prepare for this range of workforce needs. Industry needs to continue to partner with educators to articulate evolving workforce needs. Academic institutions need to be responsive and agile in building programs aligned with those needs. Policymakers should support infrastructure that allows for those workforce needs to be met. In conclusion, projected quantum industry workforce needs call for both depth and breath of expertise, which requires building educational pathways that invite a wide range of individuals into the quantum ecosystem.

\acknowledgments
Research was sponsored by the Army Research Office and was accomplished under Award Number: W911NF-24-1-0132, and NSF Grant Nos. PHY-2333073 and PHY-2333074.

\bibliography{referencesperc2025} 

\end{document}